# Optical frequency comb generation from aluminum nitride micro-ring resonator

Hojoong Jung, Chi Xiong, King Y. Fong, Xufeng Zhang, and Hong X. Tang[*]

*Department of Electrical Engineering, Yale University, New Haven, Connecticut 06511, USA*
*Corresponding author: hong.tang@yale.edu

Aluminum nitride is an appealing nonlinear optical material for on-chip wavelength conversion. Here we report optical frequency comb generation from high quality factor aluminum nitride micro-ring resonators integrated on silicon substrates. By engineering the waveguide structure to achieve near-zero dispersion at telecommunication wavelengths and optimizing the phase matching for four-wave mixing, frequency combs are generated with a single wavelength continuous-wave pump laser. The Kerr coefficient ($n_2$) of aluminum nitride is further extracted from our experimental results.

Optical frequency combs [1,2] can be generated by cascaded four-wave mixing (FWM) in materials with Kerr nonlinearity, traditionally in nonlinear optical fibers pumped by ultrafast pulsed lasers [3]. Recently, by utilizing the field enhancement effect in high quality ($Q$) factor cavities, comb generations using continuous wave (CW) laser have been demonstrated in integrated micro-scale resonators in various materials, such as silica toroids [4], doped silica glass ring resonators [5], silicon nitride (SiN) micro-ring resonators [6-8], crystalline $CaF_2$ and $MgF_2$ micro-toroids [9-11] and etc. Due to their compact size and low operating power, frequency comb generation based on micro-resonators show great potential in applications ranging from on-chip frequency references [12] to high-speed telecommunications [13].

Using aluminum nitride (AlN) as a nonlinear optical material has attracted great attention recently because of its significant second-order optical nonlinearity [14]. The high thermal conductivity of AlN (285 W/m·K), which is approximately two order of magnitude larger than that of SiN (3.2 W/m·K) [15], greatly improves the heat dissipation and thus enhances the optical devices' power handling capability. This is especially important for nonlinear optical devices which require high power to achieve strong nonlinear effect. Unlike the Si-based materials, AlN has a non-centrosymmetric crystal structure that enables both second ($\chi^{(2)}$) and third order ($\chi^{(3)}$) optical nonlinearities, making it a promising candidate for robust and high-power on-chip nonlinear optical devices. Recently, we have demonstrated the applications using AlN's $\chi^{(2)}$ nonlinearity in integrated photonic devices [16]. However, applications based on the third-order (Kerr) nonlinearity of AlN have not yet been reported. In this paper, we report the $\chi^{(3)}$-induced optical frequency comb generation in AlN micro-ring resonators and extract an approximate value of the Kerr coefficient $n_2$ of AlN based on our simulation and experimental results.

To observe the frequency comb generation, near zero group velocity dispersion (GVD) is required in the waveguide in order to satisfy the phase matching condition [7]. There are two main contributions to the dispersion of the AlN waveguide: the material dispersion, which is dominant at shorter wavelengths, and the geometric dispersion determined by the waveguide structure, which becomes more important at longer wavelengths. By choosing proper waveguide widths, near zero dispersion can be achieved at 1550 nm wavelength. Figure 1(a) shows the numerically calculated dispersion relations of AlN as a function of the waveguide width for the fundamental TE-like mode. Here, we fix the AlN waveguide height at 650 nm and vary the waveguide width from 2.5 μm to 3.5 μm. The insets show modal power profiles at 1.5 μm wavelength for different waveguide width. In the simulation we take into account the angled sidewall of the waveguide caused by the dry etching process.

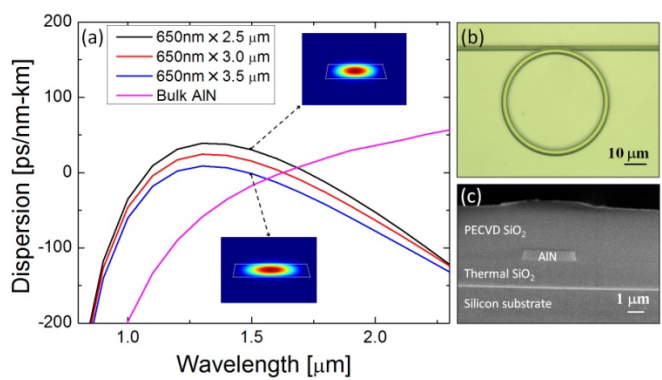

Fig. 1. (a) Simulated dispersion curve of AlN waveguide for the fundamental TE-like mode as a function of the waveguide width. The insets are modal power profiles for 2.5 μm waveguide width (top) and 3.5 μm width (bottom) at 1.5 μm wavelength. (b) Optical micrograph of the micro-ring resonator used for optical parametric oscillation. (c) Scanning electron micrograph of the cross section of the fabricated AlN waveguide.

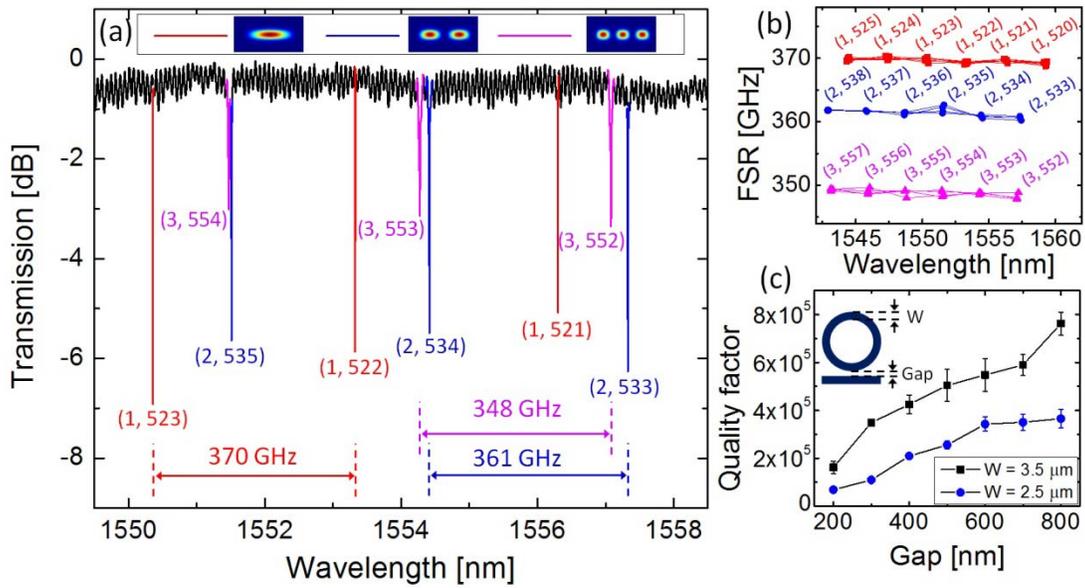

Fig. 2. (a) Transmission spectrum of a micro-ring resonator with 60 μm radius, 3.5 μm waveguide width and 500 nm gap. The modal power profile (top), FSRs, and azimuthal mode numbers of the first order mode (red), the second order mode (blue) and the third order mode (magenta) are obtained. (b) FSRs and group indices are extracted from other devices with different coupling gaps and 60 μm radius, 3.5 μm waveguide width. Numbers at each data point is the azimuthal mode number of that resonance. (c) The dependence of $Q$ factors on the gap of micro-ring resonators with 3.5 μm and 2.5 μm waveguide width. The intrinsic $Q$ factors of resonators with 3.5 μm and 2.5 μm width are 800,000 and 600,000, respectively.

Figure 1(b) is the top view of the micro-ring resonator coupled to a bus waveguide. Figure 1(c) shows a scanning electron micrograph image of the cross section of the fabricated AlN waveguide optimized for FWM. The waveguide has a height of 650 nm, a designed top width of 3.5 μm, and a slanted sidewall induced by dry etching. The bottom cladding is a layer of 2-μm-thick thermally grown $SiO_2$, and the top cladding is a 3.5-μm-thick PECVD $SiO_2$. More details of the lithography processes can be found in [14]. After the fabrication, the chip is annealed in $O_2$ atmosphere at 1000 °C for 30 hours to improve the PECVD $SiO_2$ quality that increases the quality factor of the AlN micro-ring three times higher.

The micro-ring resonator is coupled to a single mode bus waveguide with a width of 600 nm, which is adiabatically tapered down to 200 nm over a length of 500 μm near the end of the chip to create an inverted taper coupler [17]. We polish the chip facet and manage to control the tapering length with less than 10% error and the waveguide-facet angle with less than 1° error. We use a single mode fiber with 4-μm mode-field-diameter (MFD) to couple light into and out of the chip and achieve 40% fiber-to-chip coupling efficiency per facet.

The waveguide widths we choose for near-zero dispersion support multiple transverse modes at ~1550 nm wavelengths. As a result, higher order modes are observed which have different free spectral ranges (FSR), $Q$ factors and extinction ratios. In a micro-ring resonator with 60 μm radius, 3.5 μm waveguide width and 500 nm coupling gap, there are three resonant modes, the fundamental mode (red), the second order mode (blue) and the third order mode (magenta) [Fig. 2(a)]. Their FSRs are 370 GHz, 361 GHz and 348 GHz, respectively, and almost constant in this wavelength region [Fig. 2 (b)]. Using the FSRs, we can assign the azimuthal mode number $m$ of each modes. Figure 2 (c) shows the $Q$ factors of fundamental mode with different coupling gaps for 3.5 μm and 2.5 μm waveguide width. The critical coupling is achieved near 300 nm gap for 3.5 μm width and 500 nm gap for 2.5 μm width due to the different modal profile. Their estimated intrinsic $Q$ factors are 800,000 and 600,000 respectively.

For comb generation, a tunable diode laser is used in conjunction with a 3 W erbium-doped fiber amplifier (EDFA) as the pump to the input coupler. The output of the device is sent to an optical spectrum analyzer (OSA). We slowly scan the wavelength from shorter to longer wavelengths and achieve the "thermal lock" on one of the thermally shifted resonances [18]. Figure 3 shows a generated comb measured by the OSA from a micro-ring resonator with a radius of 60 μm and waveguide width of 3.5 μm. The pump power in the coupling waveguide is 500 mW and the loaded optical $Q$ factor is 600,000. The frequency comb consists of 70 peaks spanning a wavelength range of 200 nm with a 370 GHz spacing which is the FSR of the fundamental mode.

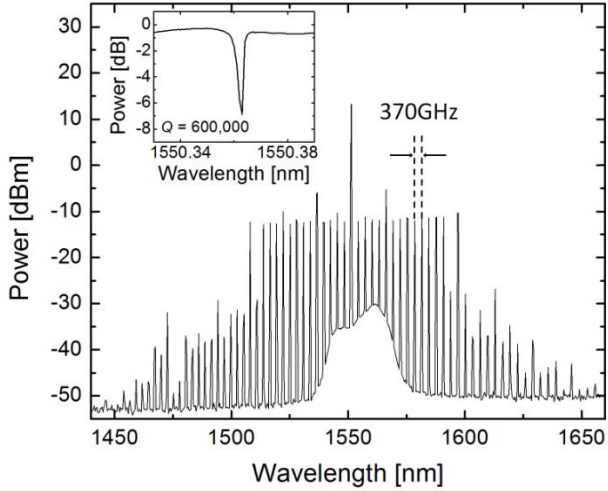

Fig. 3. Frequency comb generated from micro-ring resonators with 60 μm radius, 3.5 μm waveguide width and 500 mW pump power in the coupling waveguide. Inset is the resonance spectrum from a device with $Q$ factor of 600,000 that generated the comb.

We observe comb generations from micro-ring resonators with various radii. Comb generation in Fig. 3 is from a 60 μm-radius micro-ring, and Fig. 4 is from a 50 μm-radius micro-ring resonator. Figure 4 shows the comb spectra generated under different detuning conditions of the pump laser from the resonance. When the laser is tuned far away from the resonance such that the power in the ring just reaches the threshold power [Fig. 4(a)], we observe six generated peaks (aside from the pump peak) in the spectrum with a spacing equaling to six times of the FSR. As the pump laser is tuned into the resonance and the circulating power increases, two more wavelengths are generated in between these peaks [Fig. 4(b)]. As the pump is further tuned into the resonance, a comb of wavelengths spaced by one FSR (435 GHz) is observed [Fig. 4(c)].

The threshold power for comb generation with a waveguide width of 3.5 μm is measured to be approximately 210 mW in the waveguide. We can estimate the Kerr nonlinear coefficient $n_2$ by using the equation [19]:

$$n_2 = \frac{\omega_0^2 Q_0^{-2}(1+K)^2 + (\Delta\omega/2)^2}{P_t^{Kerr} \omega_0 \Delta\omega} \times \frac{C(\Gamma)\pi^2 R n_{eff}^2 A_{eff}}{2\lambda_0} \cdot \frac{(K+1)^2}{Q_0 K} \quad (1)$$

where $\omega_0$ is the resonant frequency, $\lambda_0$ is the corresponding resonant wavelength and $\Delta\omega = 2\omega_p - \omega_s - \omega_I$ is the frequency mismatch with $\omega_p$, $\omega_s$ and $\omega_I$ are the pump, signal and idler frequency, respectively. $P_t^{Kerr}$ is the threshold power, $R$ is the ring radius, $K = Q_0/Q_{ex}$, with $Q_0$ and $Q_{ex}$ the intrinsic and external $Q$ factors, respectively. $n_{eff}$ is the effective index, $A_{eff}$ is the effective mode area and $C(\Gamma)$ is a correction factor ranging from 1 to 2 taking into account the circulating power reduction due to the clockwise and counterclockwise modes [20]. The effective mode area ($A_{eff} = 1.34$ μm$^2$) and effective index ($n_{eff} = 1.96$) are obtained from finite element method simulations. Intrinsic $Q$ factor ($Q_0 = 800,000$), $K$ (0.3) and the frequency detuning ($\Delta\omega = 3 \pm 1.5$ GHz) are extracted from the transmission spectrum. Using those values, we estimate the Kerr coefficient of AlN for TE-like mode in (001) plane to be $n_2 = (2.3 \pm 1.5) \times 10^{-15}$ cm$^2$/W, which is comparable to that of SiN [6].

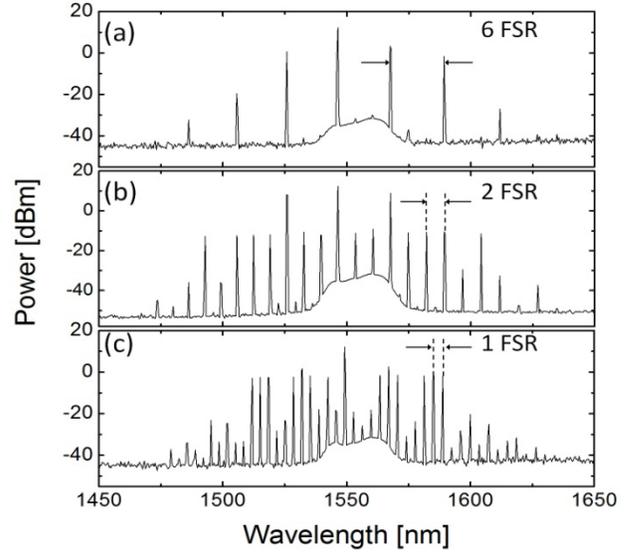

Fig. 4 The evolution of the frequency comb as the wavelength is gradually tuned to the resonant wavelength from a micro-ring with a radius of 50 μm and waveguide width of 2.5 μm and 700 mW pump power in the coupling waveguide. When the wavelength approaches to the resonance, peaks start to appear with six FSR spacing [(a)], which decreases to two [(b)] and eventually one [(c)] FSR when the wavelength gets even closer to the resonance.

In conclusion, we observed optical frequency comb generation from high-$Q$ micro-ring resonators made from AlN. We show that besides its known $\chi^{(2)}$ nonlinearity, AlN also possesses strong $\chi^{(3)}$ nonlinearity for producing efficient four wave mixing and generating optical parametric oscillation and frequency combs. We investigate the power dependence of the comb generation and estimate the Kerr coefficient of AlN based on the experiment and simulation data. The comb generation threshold can be further reduced if the optical $Q$-factor can be increased by improving the dry etching recipe and the sidewall roughness. We also note that there is also room for further improvement by reducing the coupling loss of inverted tapers. With both strong Kerr nonlinearity and electro-optic properties, AlN is promising for building electrically-tunable and high-speed-switchable frequency comb devices, which have applications in optical clocks, metrology, and chip-scale high speed communications [21].

This work was supported by a Defense Advanced Research Projects Agency (DARPA) grant managed by Dr. J. R. Abo-Shaeer. H.X.T. acknowledges support from a Packard Fellowship in Science and Engineering and a National Science Foundation CAREER award. Facilities used were supported by Yale Institute for Nanoscience and Quantum Engineering and NSF MRSEC DMR 1119826. The authors thank Michael Power and Dr. Michael Rooks for assistance in device fabrication.

References.

1. T. J. Kippenberg, R. Holzwarth, and S. A. Diddams, Science **332**, 555 (2011).
2. S. T. Cundiff and Jun Ye, Rev. Mod. Phys. **75**, 325 (2003)
3. S. A. Diddams, D. J. Jones, J. Ye, S. T. Cundiff, J. L. Hall, J. K. Ranka, R. S. Windeler R. Holzwarth, T. Udem, and T. W. Hänsch, Phys. Rew. Lett. **84**, 5102 (2000).
4. P. Del'Haye, A. Schliesser, O. Arcizet, T. Wilken, R. Holzwarth, and T. J. Kippenberg, Nature **450**, 1214 (2007).
5. L. Razzari, D. Duchesne, M. Ferrera, R. Morandotti, S. Chu, B. E. Little, and D. J. Moss, Nat. Photon. **4**, 41 (2010).
6. J. S. Levy, A. Gondarenko, M. A. Foster, A. C. Turner-Foster, A. L. Gaeta, and M. Lipson, Nat. Photon. **4**, 37 (2010).
7. Y. Okawachi, K. Saha, J. S. Levy, Y. H. Wen, M. Lipson, and A. L. Gaeta, Opt. Lett. **36**, 3398 (2011).
8. A. R. Johnson, Y. Okawachi, J. S. Levy, J. Cardenas, K. Saha, M. Lipson, and A. L. Gaeta, Opt. Lett. **37**, 875 (2012).
9. I. S. Grudinin, N. Yu, and L. Maleki, Opt. Lett. **34**, 878 (2009).
10. A. A. Savchenkov, A. B. Matsko, V. S. Ilchenko, I. Solomatine, D. Seidel, and L. Maleki, Phys. Rev. Lett. **101**, 093902 (2008).
11. C.Y. Wang, T. Herr, P. Del'Haye, A. Schliesser, J. Hofer, R. Holzwarth, T.W. Hänsch, N. Picque, and T. J. Kippenberg, Nat. Comm. **4**, 1345 (2013).
12. Th. Udem, R. Holzwarth, and T. W. Hänsch, Nature **416,** 233 (2002).
13. D. Hillerkuss, R. Schmogrow, T. Schellinger, M. Jordan, M. Winter, G. Huber, T. Vallaitis, R. Bonk, P. Kleinow, F. Frey, M. Roeger, S. Koenig, A. Ludwig, A. Marculescu, J. Li, M. Hoh, M. Dreschmann, J. Meyer, S. Ben Ezra, N. Narkis, B. Nebendahl, F. Parmigiani, P. Petropoulos, B. Resan, A. Oehler, K. Weingarten, T. Ellermeyer, J. Lutz, M. Moeller, M. Huebner, J. Becker, C. Koos, W. Freude, and J. Leuthold, Nat. Photon. **5**, 364 (2011).
14. C. Xiong, W. H. P. Pernice1, X. Sun, C. Schuck, K. Y. Fong, and H. X. Tang, New J. Phys. **14** 095014 (2012).
15. C. H. Mastrangelo, Y-C Tai, and R. S. Muller, Sensors Actuators A **23** 856 (1990)
16. C. Xiong, W. H. P. Pernice, and H. X. Tang, Nano Lett. **12**, 3562 (2012)
17. G. Ren, S. Chen, Y. Cheng, and Y. Zhai, Opt. Comm. **284**, 4782 (2011).
18. T. Carmon, L. Yang, and K. J. Vahala, Opt. Express **12**, 4742 (2004).
19. T. J. Kippenberg, S. M. Spillane, and K. J. Vahala, Phys. Rev. Lett. **93**, 083904 (2004).
20. T. J. Kippenberg, S. M. Spillane, and K. J. Vahala, Opt. Lett. **27**, 1669 (2002).
21. N. R. Newbury, Nat. Photon. **5**, 186 (2011).